# Magnetism and in-gap states of 3*d* transition metal atoms on superconducting Re

Lucas Schneider[1], Manuel Steinbrecher[1,‡], Levente Rózsa[1], Juba Bouaziz[2], Krisztián Palotás[3,4,5,6], Manuel dos Santos Dias[2], Samir Lounis[2], Jens Wiebe[1,*], and Roland Wiesendanger[1]

[1]Department of Physics, University of Hamburg, D-20355 Hamburg, Germany.

[2]Peter Grünberg Institute and Institute for Advanced Simulation, Forschungszentrum Jülich & JARA, D-52428 Jülich, Germany.

[3]Department of Theoretical Physics, Budapest University of Technology and Economics, H-1111 Budapest, Hungary.

[4]Institute for Solid State Physics and Optics, Wigner Research Center for Physics, Hungarian Academy of Sciences, P. O. Box 49, H-1525 Budapest, Hungary.

[5]MTA-SZTE Reaction Kinetics and Surface Chemistry Research Group, University of Szeged, H-6720 Szeged, Hungary.

[6]Institute of Physics, Slovak Academy of Sciences, SK-84511 Bratislava, Slovakia.

[‡]Present address: Institute for Molecules and Materials (IMM), Radboud University, Nijmegen, The Netherlands.

[*]E-mail: jwiebe@physnet.uni-hamburg.de

**Magnetic atoms on heavy-element superconducting substrates are potential building blocks for realizing topological superconductivity in one- and two-dimensional atomic arrays. Their localized magnetic moments induce so-called Yu-Shiba-Rusinov (YSR) states inside the energy gap of the substrate. In the dilute limit, where the electronic states of the array atoms are only weakly coupled, proximity of the YSR states to the Fermi energy is essential for the formation of topological superconductivity in the band of YSR states. Here, we reveal via scanning tunnel spectroscopy and *ab initio* calculations of a series of 3*d* transition metal atoms (Mn, Fe, Co) adsorbed on the heavy-element superconductor Re that the increase of the Kondo coupling and sign change in magnetic anisotropy with *d*-state filling is accompanied by a shift of the YSR states through the energy gap of the substrate and a crossing of the Fermi level. The uncovered systematic trends enable the identification of the most promising candidates for the realization of topological superconductivity in arrays of similar systems.**

Recently, theoretical predictions and experimental indications of topological superconductivity in one-[1–15] and two-dimensional[16–20] arrays of magnetic atoms coupled to bulk superconductors have triggered renewed interest in the building blocks of such systems, i.e. single magnetic atoms coupled to bulk superconductors[21–30]. Magnetic atoms locally induce pairs of bound state resonances inside the gap of their host superconductor, known as Yu-Shiba-Rusinov (YSR) states[31–35]. One route towards topological superconductivity is to weakly couple these states in so-called YSR bands, which can be realized by a dilute array of magnetic atoms on the surface of a superconductor[17,36]. In order to enter the topologically nontrivial phase, the YSR band has to cross the Fermi level $E_F$. This requires the width in energy and the separation from $E_F$ of the YSR states of the atomic building blocks to be sufficiently small.

In a classical picture, the energy position $E_{YSR}$ of the YSR state with respect to $E_F$ depends mainly on the *s-d-* or Kondo exchange coupling $J_K$ of the local spin of the atom to the substrate conduction electrons.[34] For low $J_K$, the peaks merge with the coherence peaks at the gap edge $\Delta$, whereas for higher $J_K$, they shift towards $E_F$ and eventually cross it, commonly referred to as a quantum phase transition[37]. For even higher $J_K$, the in-gap states merge with the coherence peaks again. However, different effects complicate this picture calling for detailed experimental investigations. The YSR state is accompanied by the formation of the Kondo state[23,30]. Orbital effects[22,27,29] and magnetic anisotropy[25] can lead to its multiplicity, and exchange interactions between the atoms may further split or shift the YSR states[22,38–41]. It is known since the 1960s that there is a systematic trend in the evolution of the Kondo screening, which is intimately connected to $J_K$, within the series of 3*d* transition metal atoms doped into the bulk or onto the surface of various host metals. The Kondo temperature decreases for each unpaired *d* electron in the band, showing a minimum in the middle of the series for Mn and monotonically increasing towards lower (Cr, V and Ti) and higher (Fe, Co and Ni) occupations[42–44]. Since this systematic behavior is driven by the successive occupation of the *d* orbitals, it suggests similar trends for other properties related to $J_K$, such as the energies of YSR states. However, an experimental investigation of the YSR binding energy in a series of 3*d* transition metal atoms adsorbed on the same superconducting substrate has been lacking so far. Such a classification is strongly desired regarding the design of topological superconductivity in dilute arrays of atoms on elemental superconductors.

Here, we report on a combined experimental and *ab initio* investigation for the heavy-element superconductor Re as a substrate ($T_c \approx 1.6$ K), for which strong indications of topological superconductivity have been previously reported in densely packed iron chains.[15] We show that the YSR state energy is shifting through the gap by increasing the 3*d*-orbital occupation going from the middle towards the right end of the 3*d* transition metal series, which correlates with a simultaneous increase in the Kondo coupling, and a transition from easy-axis to easy-plane magnetic anisotropy as revealed by spin-excitation spectroscopy.

## Results

**Topographic signatures**
The preparation of an atomically clean (0001) surface of a Re single crystal and subsequent successive deposition of Mn, Fe, and Co onto the cold substrate ($T < 10$ K, see Methods) usually results in a statistical distribution of atoms on the two possible hollow adsorption sites, face-centered cubic (fcc) and hexagonal close packed (hcp). Figures 1a and b show constant-current images of samples where either Mn and Fe (a) or Co and Fe (b) have been deposited. Prior to these co-deposition experiments, all species were individually deposited and investigated which enabled us to discern them. The atoms can be distinguished by their apparent height and their adsorption site with respect to the atomic lattice of the Re(0001) surface (Fig. 1a) determined from manipulated atom images[45]. The adsorption sites of Mn and Co have been determined by their relative positions with respect to that of Fe$_{fcc}$ and Fe$_{hcp}$ atoms which have been characterized in a previous work.[15] Fe atoms have a fcc-hcp ratio of about 1:4 after deposition, appear with heights of 65 pm for Fe$_{fcc}$ and 75 pm for Fe$_{hcp}$, and can be manipulated between the two sites by atom manipulation (see Methods). Mn atoms adsorb only on the fcc hollow sites, appear with a height of 90 pm and cannot be manipulated to the other site. Co atoms occupy only hcp hollow sites after deposition, but can be manipulated to the fcc site and *vice versa*. They appear with heights of 95 pm for Co$_{fcc}$ and 85 pm for Co$_{hcp}$.

**Kondo resonances and spin excitations**
In Figs. 1c-g, we start by studying the spectroscopic signatures of the different species on the (0001) surface of Re in the normal metallic state in a larger energy window around $E_F$. Superconductivity was quenched by applying an external magnetic field of $B_z = 50$ mT, higher than the critical field of bulk Re ($B_c = 20$ mT)[46], perpendicular to the surface. The Mn$_{fcc}$ atom in Fig. 1c features a step-shaped excitation around ±1 meV on a largely flat background, which is assigned to a spin excitation as discussed below. Due to the absence of a resonance around the Fermi energy, we conclude that the Kondo energy scale $k_B T_K$ is considerably smaller than the experimental thermal energy of 0.025 meV. On the other hand, the Co$_{fcc}$ atom in Fig. 1f reveals a broad resonance, which indicates a relatively strong Kondo coupling of its magnetic moment. Using the fit to a Frota function (dashed line, see Supplementary Note 3), a correspondingly large Kondo energy of $k_B T_K \approx 1.03$ meV can be approximated via Wilson's[47] definition $k_B T_K = 0.27 \Gamma_{FWHM}$ from the half width at half maximum $\Gamma_{FWHM}$. The Co$_{hcp}$ atom in (g) shows neither a spin excitation nor a Kondo resonance in a range of ±20 meV around $E_F$, signaling a vast quenching of its magnetic moment. The Fe$_{fcc}$ (d) and the Fe$_{hcp}$ atoms (e) reveal both kinds of features, i.e. symmetric steps around ±1 meV (see arrows) and Fano resonances at $E_F$, indicating an interplay of spin excitations and relatively weak Kondo coupling.[48] Comparably small Kondo energies of $k_B T_K \approx 0.18$ meV (Fe$_{fcc}$) and 0.10 meV (Fe$_{hcp}$) are consistently estimated from the fits to Frota functions (dashed curves, see Supplementary Note 3). Note that there is an additional broad resonance for Fe$_{hcp}$ which points towards a second Kondo screening channel with a larger Kondo temperature comparable to that of Co$_{fcc}$.

**Magnetic moments and anisotropies**
In Fig. 2, we investigate the magnetic-field dependence of the different spectroscopic features. The excitation observed on Mn$_{fcc}$, see Fig. 2a, linearly shifts as a function of $B_z$ reminiscent of the spin excitation of an atom with a uniaxial out-of-plane magnetic anisotropy.[49] For Fe$_{fcc}$, see Fig. 2b, the resonance and spin excitation evolve into two step-like features (each above and below $E_F$), the former almost linearly increasing and the latter decreasing in energy with a crossing at $B_z \approx 6$ T. For $B_z \approx 11.5$ T the latter reaches $E_F$ accompanied by the reappearance of a zero-bias resonance (see Supplementary Figure 2). This behavior is very reminiscent of the $B_z$ evolution of the features seen for an atom with easy-plane magnetic anisotropy[50]. For Fe$_{hcp}$, see Fig. 2c, the behavior is qualitatively similar (see also Supplementary Figure 4), but heavily obscured by a strongly asymmetric Fano line shape of the zero-bias resonance, the weakness of the excitation feature (arrows) and the broad background resonance. Finally, the Kondo resonance seen on the Co$_{fcc}$, see Fig. 2d, does not show any change up to 8 T, as expected from a strongly Kondo-coupled impurity.[51] The Co$_{hcp}$ spectrum remains featureless in external magnetic field.

In order to further support these conclusions on the magnetic anisotropies and Kondo couplings, and to quantify these parameters, we simulate the tunnel spectroscopy data using models based on the effective spin Hamiltonian

$$\hat{H} = \frac{K}{S^2} \cdot \hat{S}_z^2 + g \cdot \mu_B \cdot B_z \cdot \hat{S}_z, \qquad (1)$$

which includes a uniaxial anisotropy term with the coefficient of the magnetic anisotropy $K$, the spin quantum number $S$, and the $z$-component of the spin operator $\hat{S}_z$, and a Zeeman term with the atom's *g*-factor, the Bohr magneton $\mu_B$, and the strength of the magnetic field applied in the *z*- direction perpendicular to the substrate plane $B_z$. The spin value $S$ was determined by calculating the total magnetic moment of the atoms $\mu$ from *ab initio* electronic structure methods, and using the relation $\mu = g S \mu_B$ with the assumption $g = 2$. The geometry was optimized and the preferred adsorption sites were determined by using the Vienna Ab-initio Simulation Package (VASP, see Methods and Supplementary Note 4)[52–54]. For calculating the magnetic moments, besides the VASP package, two implementations of the Korringa–Kohn–Rostoker (KKR) Green function method based on an embedding scheme have been applied: the code developed in Budapest (KKRBp)[55,56] and the program developed in Jülich (KKRJ)[57–59] (see Methods). The VASP results are summarized in Table 1; the results of the other methods are given in Supplementary Note 5. The energy differences between the adsorption sites fit well to our experimental findings of the preferred sites. The magnetic moments of the atoms show the usual decrease from the middle towards the end of the series (Mn: $\sim 4\mu_B$; Fe: $\sim 3\mu_B$; Co: $\sim 2\mu_B$), but with an overall reduction by $1\mu_B$ as compared to the values of the free-standing atoms calculated from Hund's rules for 5, 4, and 3 unpaired *d* electrons, and a consistently smaller moment for the hcp with respect to the fcc species. The decreased magnetic moments of the atoms may be attributed to the delocalization of the *d* electrons due to the hybridization with the metallic substrate, which is stronger for the hcp site due to a larger relaxation towards the surface (Supplementary Note 4). The magnetic moment of $Co_{hcp}$ decreases with respect to that of $Co_{fcc}$ and almost vanishes within KKRBp, indicating the strong tendency towards a complete quenching of its magnetism, which can be rationalized within the Stoner picture (Supplementary Note 5). The resulting approximate values of the spin quantum number $S$ used within the effective spin model are given in Table 1. Note, that the full physics of 3*d* transition metal atoms adsorbed on the metallic surface of a 5*d* transition metal is correctly described within the concept of a Hund's impurity[48], where the assumption of a discrete integer or half-integer spin is no longer valid due to strong charge fluctuations. However, as was shown before[48], the magnetic-field dependence of the spin excitations of a Hund's impurity are, in many cases, modeled surprisingly well by the perturbation theory model[60] using the effective spin Hamiltonian. Therefore, we use it here in order to rationalize the excitations and zero-bias resonances, as well as to quantify the magnetic anisotropies and Kondo couplings.

The magnetic-field dependence of the simulated tunnel spectra using the perturbation-theory model[60] excellently reproduce the experimental data of $Mn_{fcc}$ and $Fe_{fcc}$ (see Supplementary Figure 3 for a comparison of experimental data and simulation). The fitted values for magnetic anisotropy ($K$) and product of Kondo coupling with the density of states of the substrate at the Fermi level ($J_K \rho_F$) are $K = -1.56$ meV and $J_K \rho_F = -0.02$ for $Mn_{fcc}$, i.e. a uniaxial easy-axis anisotropy with negligible Kondo coupling, and $K = +1.55$ meV and $J_K \rho_F = -0.45$ for $Fe_{fcc}$, i.e. an easy-plane anisotropy and weak Kondo coupling (remaining parameters in Supplementary Note 3). The corresponding level schemes are shown in Figs. 2e and f. In particular, the zero-bias resonance and excitation step observed on the $Fe_{fcc}$ atom can now be assigned to the Kondo state involving the degenerate $S_z = \pm 1/2$ spin levels, and degenerate spin excitations into the $S_z = \pm 3/2$ levels, respectively. They are evolving into two spin excitations (-1/2 to +1/2 and -3/2), the former increasing and the latter decreasing in energy with $B_z$ (see arrows in Fig. 2f), which finally leads to the reappearance of the Kondo peak when the $S_z = -1/2$ and $S_z = -3/2$ levels cross around $B_z = 11.5$ T (Supplementary Note 2). A similar easy-plane anisotropy level scheme is consistent with the experimental data of $Fe_{hcp}$, see Fig. 2g. However, due to the strongly asymmetric Fano line shape of the zero-bias resonance and the broad resonance of the background, the spectra cannot be fitted with the used perturbation-theory code, that does not yet include interference effects from different tunnel paths. Instead, we used a superposition of magnetic-field dependent Frota functions and steps resulting in $K = +0.79$ meV (Supplementary Note 3). For $Co_{fcc}$ the indifference of the broad resonance to the magnetic field up to $B_z = 8$ T cannot be reconciled with the perturbation theory simulations assuming the calculated effective spin of $S = 1$. In contrast, it would be more consistent with a $S = 1/2$ impurity system in the strong Kondo-coupling regime, where the Kondo resonance is relative insensitive to moderate magnetic fields, as shown by numerical renormalization group calculations[51]. Our experimental data therefore suggests the level scheme in Fig. 2h. Finally, for $Co_{hcp}$ the absence of any spectroscopic features around $E_F$ is consistent with the quenching of the magnetic moment obtained from the KKRBp *ab initio* calculations as explained within the Stoner model.

The extracted experimental magnetic anisotropies are summarized in Table 1 together with the experimental Kondo energies. As also shown in the table, the experimentally extracted parameters are qualitatively reproduced by the magnetic anisotropy values calculated from the VASP method (see Methods) and by the Kondo couplings calculated from time-dependent density functional theory (TD-DFT) within the KKRJ method (Supplementary Note 6). The combined data reveals a systematic decrease in the magnetic moment and an increase in the Kondo coupling going from the center of the 3*d* transition metal series (Mn) towards the end (Co) and from fcc to hcp adsorption site. The determined Kondo energy scales thereby vary from values that are much smaller than the energy gap of the substrate $\Delta_{Re} \approx 0.255$ meV (for $Mn_{fcc}$), over values comparable to $\Delta_{Re}$ (for $Fe_{fcc}$ and $Fe_{hcp}$), to values that are much larger than $\Delta_{Re}$ (for $Co_{fcc}$). This trend is accompanied, for the particular case of adsorption on the Re(0001) substrate, by a transition in the magnetic anisotropy from easy-axis to easy-plane, where the absolute values of the magnetic anisotropy are several times larger than $\Delta_{Re}$. In the following, we will experimentally investigate how these magnetic properties reflect in the YSR state binding energies of all species when the substrate is driven into the superconducting state ($B_z = 0$ T).

**Yu-Shiba-Rusinov states**

Tunnel spectroscopy performed on the Re substrate reveals the typical energy gap of $\Delta_{Re}$ with coherence peaks at the gap edges (Fig. 3a,c,e, gray dashed lines). Spectra taken on top of $Mn_{fcc}$ (Fig. 3a, green) show a tiny shift of spectral weight from the coherence peaks into the band gap. The corresponding spectroscopic features get more obvious in the normalized spectrum (Fig. 3b) exhibiting two peaks inside the gap, energetically symmetric around $E_F$, which we assign to the YSR states of $Mn_{fcc}$. The YSR states almost merge with the original coherence peaks. Considering the small Kondo coupling discussed above, we conclude that the energy shift of the $Mn_{fcc}$ YSR state from the gap edge is very small, i.e. $E_{YSR}^{Mn_{fcc}} \approx +0.23$ meV with respect to $E_F$. In contrast, for both types of Fe atoms, the resonances are well inside the gap, see Figs. 3c and d, which we assign to moderately bound YSR states. For the $Fe_{fcc}$ atom the energy is around $E_{YSR}^{Fe_{fcc}} \approx +0.12$ meV. For the $Fe_{hcp}$ atom the asymmetric shape of the tunnel spectra consists of a pair of YSR resonances at the position of roughly $E_{YSR}^{Fe_{hcp}} \approx +0.04$ meV compared to $E_F$, as was investigated in Ref. [15]. This is consistent with the observation of a larger Kondo coupling for Fe as compared to Mn. Interestingly, the narrow zero-bias features of both Fe species associated with the weak-coupling Kondo resonances shown in Fig. 1d,e are completely suppressed in the superconducting state of the Re substrate (see Supplementary Note 1). For the $Co_{fcc}$ magnetic impurity the YSR states are again almost merging with the coherence peaks in Figs. 3e and f. Since it was found that this atom possesses the strongest Kondo coupling, we suppose that the YSR states have already crossed $E_F$, and are located at $E_{YSR}^{Co_{fcc}} \approx -0.24$ meV, in contrast to the positive value assigned to the $Mn_{fcc}$ atom. Finally, the $Co_{hcp}$ impurity does not show any signature of in-gap states, which is consistent with the full quenching of its magnetic moment.

**Discussion**

The experimentally measured YSR binding energies are summarized in Table 1 together with the magnetic moments, anisotropies, and Kondo energies. We see that going from the center of the 3*d* transition metal series (Mn) towards the end (Co), and from fcc to hcp adsorption site, the decrease in $\mu$ and increase in $k_B T_K$ correlates with a systematic shift of the YSR state from the coherence peak at one gap edge across $E_F$ towards the coherence peak on the other side of the gap. This evolution is expected from the classical picture[31–35] in connection with the well-known systematics of the Kondo coupling of impurities from the 3*d* series[42–44], but is proven here experimentally for the first time. Interestingly, the observed simultaneous transition of the magnetic anisotropy from out-of-plane to easy-plane, which usually complicates the situation by inducing additional energetic shifts of the YSR states,[25] does not interfere with the above trend imposed by the Kondo coupling. In contrast, it seems to go hand in hand with this trend by quenching the Kondo screening in the out-of-plane magnetic anisotropy case and supporting the Kondo screening in the easy-plane case.[48]

In consequence of this systematics, the magnetic state of $Fe_{hcp}$ is closest to the situation where the YSR state just crosses $E_F$. This result has interesting implications considering the theoretical suggestions for the emergence of topological superconductivity in dilute arrays of weakly coupled magnetic atoms on the surface of heavy-element superconductors[36], i.e. to choose magnetic impurities whose YSR states have a broad spectral weight at $E_F$. From our results, we expect that arrays of weakly coupled $Fe_{hcp}$ atoms on Re are the most promising candidates to realize one- and two-dimensional topological superconductivity. As the Dzyaloshinskii-Moriya component of the Ruderman-Kittel-Kasuya-Yosida exchange interaction is usually strong for the heavy-element substrate,[15,61] the easy-plane magnetic anisotropy of $Fe_{hcp}$ facilitates the formation of non-collinear spin states in an array. This can further support topological superconductivity.

We finally expect a similar trend in the evolution of the YSR state energy for lower occupation of the 3$d$ states, i.e. going from Mn over Cr and V to Ti. Therefore, we forecast that the YSR state of Cr or V on Re might be close to $E_F$. We also expect that a comparable systematics exists for the series of 3$d$ transition metal atoms on other heavy-element superconducting substrates. Our results therefore enable to steer the search for topological superconductivity in the platform of dilute arrays of 3$d$ transition metal atoms on heavy-element superconducting substrates.

## Methods

**Experimental procedures**

All measurements were performed in a home-built ultra-high-vacuum scanning tunneling microscope (STM) setup at $T = 0.3$ K with an optional magnetic field $B_z$ of at most 12 T applied perpendicular to the sample surface[62]. We used electrochemically edged tungsten tips that were flashed to $T = 1500$ K before inserting them into the STM. The Re(0001) crystal was cleaned by Ar ion sputtering, followed by multiple cycles of $O_2$ annealing at $T = 1530$ K and flashing to $T = 1800$ K. Mn, Fe and Co atoms were successively deposited keeping the substrate at $T < 10$ K. The bias-dependent differential conductance d$I$/d$V$ was measured using a Lock-In amplifier by modulating the bias voltage $V$ with $V_{mod}$ = 0.04-0.20 mV at a frequency of $f_{mod}$ = 4142 kHz, and at constant tip height stabilized at a bias voltage $V_{stab}$ and tunnel current $I_{stab}$ before opening the feedback loop for measurement. Note that the bias voltage is applied to the sample and zero bias corresponds to $E_F$. In order to remove any effects in the tunnel spectra measured in this way from a residual variation in the tip density of states and in order to increase the visibility of spectral features stemming from the atoms, tunnel spectra taken on the atom were normalized by division ("norm.") or subtraction ("difference") of a substrate spectrum taken with the same microtip, and are called "normalized d$I$/d$V$". Single atoms were manipulated using STM-tip-induced atom manipulation by lowering the bias voltage and increasing the setpoint current to the manipulation parameters $V$ = 1 mV and $I$ = 100 nA.

**VASP-based calculations**

During the VASP[52–54] calculations, the exchange-correlation functional was parametrized using the Perdew-Burke-Ernzerhof (PBE)[63] method within the generalized gradient (GGA) approximation. The considered system consisted of 4 layers of Re in hcp stacking along the (0001) direction with 7×7 atoms in each layer and a single 3$d$ adatom. The in-plane lattice constant was chosen to be $a$ = 2.761 Å, and the bulk interlayer distance was set to $d$ = 2.228 Å. A vacuum region of minimum 9 Å was kept between the slabs in order to minimize interactions between them. The Brillouin zone was sampled by the Gamma point only because of the large supercell size. The positions of the bottom three Re layers were kept fixed during the calculations, while the Re atoms in the top layer and the adatom were allowed to relax along the direction perpendicular to the surface. It was found that allowing the atoms to also move in the plane does not make any of the adatoms switch between fcc and hcp adsorption sites, primarily because the force acting on them always points along the out-of-plane direction due to the symmetric arrangement of the neighboring Re atoms. The anisotropy coefficients $K$ were determined from total energy differences between in-plane ($x$) and out-of-plane ($z$) orientations of the magnetic moment of the adatom, $K = E_{\text{tot}}(e_\mu \parallel e_z) - E_{\text{tot}}(e_\mu \parallel e_x)$, where $e_\mu$, $e_x$, and $e_z$ denote unit vectors along the magnetic moment direction as well as along the $x$ and $z$ axes. Spin-orbit coupling was included in the anisotropy calculations and $3 \times 3 \times 1$ $k$ points were considered in the Brillouin zone integration.

### KKRJ-based calculations

Within the KKR code developed in Jülich[57] the atomic sphere approximation (ASA) was used with an angular momentum cutoff of $l_{\text{max}} = 3$ Calculations based on the full-potential version of the code are presented in the Supplementary Material. The exchange and correlation potential was treated within the local density approximation (LDA) using the parametrization of Vosko, Wilk and Nusair[64]. The unit cell used for the slab calculations contained 20 Re layers in hcp stacking. The Re substrate was converged self-consistently using $40 \times 40$ $k$ points in the full Brillouin zone. It was found that the size of the embedding cluster used to simulate the adatom was not of the same importance as recently experienced on the Pt substrate[49] for the convergence of different quantities of interest, such as the magnetic moments or the magnetic anisotropy energy. Therefore, we considered a real-space cluster containing 47 sites, including 37 Re atoms, which was built around a vacuum site located above the Re(0001) surface for the two different adsorption sites, hcp and fcc. Mn, Fe and Co impurities were then embedded self-consistently above the Re substrate. The vertical positions of the adatoms during the calculations are reported in Supplementary Note 4.

### KKRBp-based calculations

Besides the VASP and KKRJ methods, we also performed electronic structure calculations using the KKR code developed in Budapest (KKRBp)[56]. The potential of bulk Re was determined self-consistently as a first step. Then the surface was constructed from 8 layers of Re and 4 layers of empty spheres (vacuum) sandwiched between semi-infinite bulk Re and semi-infinite vacuum. The geometrical parameters were the same as in the KKRJ calculations listed in Supplementary Note 4, with the adatoms to be embedded in the vacuum layer closest to the surface. Also similarly to the KKRJ calculations, the atomic sphere approximation (ASA) was used with an angular momentum cutoff of $l_{\text{max}} = 3$. The parametrization of the exchange-correlation potential within the Local Density Approximation (LDA) was based on the Ceperley-Alder method.[65] The energy integration was performed on a semicircle contour containing 16 points, with up to 547 $k$ points in the irreducible part of the Brillouin zone at the energy points close to the Fermi level. The self-consistent calculations for the adatoms were carried out by embedding a cluster of atoms in the layered system.[55] For the hcp adsorption site the cluster contained 128 atomic spheres (1 adatom, 61 Re and 66 vacuum), while 136 atomic spheres (1 adatom, 69 Re and 66 vacuum) were included for the fcc adsorption site. The direction of the exchange-correlation magnetic field $B_{\text{xc}}$, which should be parallel to the magnetic moment in the ground state, was oriented along the out-of-plane direction for all atoms in the cluster during the self-consistent calculations.

### Data availability

The authors declare that the main data supporting the findings of this study are available within the article and its Supplementary Information files. Extra data are available from the corresponding author upon reasonable request.

## Acknowledgements
L.S., M.S., L.R., R.W., and J.W. acknowledge funding by the Cluster of Excellence 'Advanced Imaging of Matter' (EXC 2056 - project ID 390715994) of the Deutsche Forschungsgemeinschaft (DFG). L.S. and R.W. acknowledge funding by the ERC Advanced Grant ASTONISH (No. 338802). R.W. acknowledges funding by the ERC Advanced Grant ADMIRE (No. 786020). L.R. acknowledges funding by the Alexander von Humboldt Foundation. K.P. acknowledges support from the National Research Development and Innovation Office of Hungary project nos. K115575 and FK124100, the BME-Nanotechnology FIKP grant of EMMI, and the Slovak Academy of Sciences project no. SASPRO-1239/02/01. J.B., M.d.S.D. and S.L. acknowledge computing time granted through JARA-HPC on the supercomputer JURECA at the Forschungszentrum Jülich and funding from the European Research Council (ERC) under the European Union's Horizon 2020 research and innovation program (ERC-consolidator grant 681405 – DYNASORE).

## Author contributions
L.S., M.S., R.W., and J.W. designed the experiments. L.S. and M.S. carried out the measurements. L.S. and J.W. did the analysis and the simulation of the experimental data. L.R. and K.P. performed the VASP and KKRBp-based calculations. J.B. performed the KKRJ-based and TD-DFT calculations. L.R., J.B., K.P., M.d.S.D., and S.L. analysed the *ab initio* simulations. L.S. and J.W. wrote the manuscript, to which all authors contributed via discussions and corrections.

## Competing interests.
The authors declare no competing interests.

# Tables

| | $\Delta E$ [meV] | $\mu_{\text{VASP}}$ [$\mu_B$] | $S = \dfrac{\mu_{\text{VASP}}}{2\mu_B}$ | $k_B T_K / \Delta_{\text{Re}}$ experiment | $\|J_K \rho_F\|$ calc. | $K_{\text{exp}} / \Delta_{\text{Re}}$ experim. | $K_{\text{VASP}} / \Delta_{\text{Re}}$ calculated | $E_{\text{YSR}} / \Delta_{\text{Re}}$ experiment |
|---|---|---|---|---|---|---|---|---|
| Mn$_{\text{fcc}}$ | -14.6 | 3.94 | 2 | <0.1 | 0.32 | -6.12 | -10.78 | +0.90 |
| Mn$_{\text{hcp}}$ | | 3.76 | 2 | - | 0.42 | - | -10.86 | - |
| Fe$_{\text{fcc}}$ | 139.5 | 3.08 | 3/2 | 0.71 | 0.82 | +6.08 | +10.27 | +0.47 |
| Fe$_{\text{hcp}}$ | | 2.84 | 3/2 | 0.39 | 0.91 | +3.10 | +6.04 | +0.16 |
| Co$_{\text{fcc}}$ | 229.2 | 2.02 | 1 (1/2) | 4.04 | 1.52 | - | +16.27 | -0.94 |
| Co$_{\text{hcp}}$ | | 1.54 | 1 (0) | - | 1.97 | - | +2.98 | - |

**Table 1 | Comparison of the parameters from the experiments and from the *ab initio* calculations.** Energy difference between the adsorption sites $\Delta E = E_{\text{fcc}} - E_{\text{hcp}}$ obtained from VASP calculations, total magnetic moment of the atom $\mu_{\text{VASP}}$ obtained from VASP, and the resulting effective spin values $S$ (note that, for Co, the experimental data suggests the values given in brackets). Experimental Kondo energy scales $k_B T_K$ and Kondo couplings $|J_K \rho_F|$ calculated from TD-DFT within KKRJ. Uniaxial anisotropy coefficients $K$ deduced from the fitting of the experimental spectra to the perturbation theory (Mn$_{\text{fcc}}$ and Fe$_{\text{fcc}}$) and Multi-Frota (Fe$_{\text{hcp}}$) models using the calculated effective spins $S$, in comparison to the values obtained from the VASP method. YSR state energies $E_{\text{YSR}}$ extracted from the experimental data. The energies are given with respect to the substrate energy gap of $\Delta_{\text{Re}} \approx 0.255$ meV, where meaningful.

# Figures

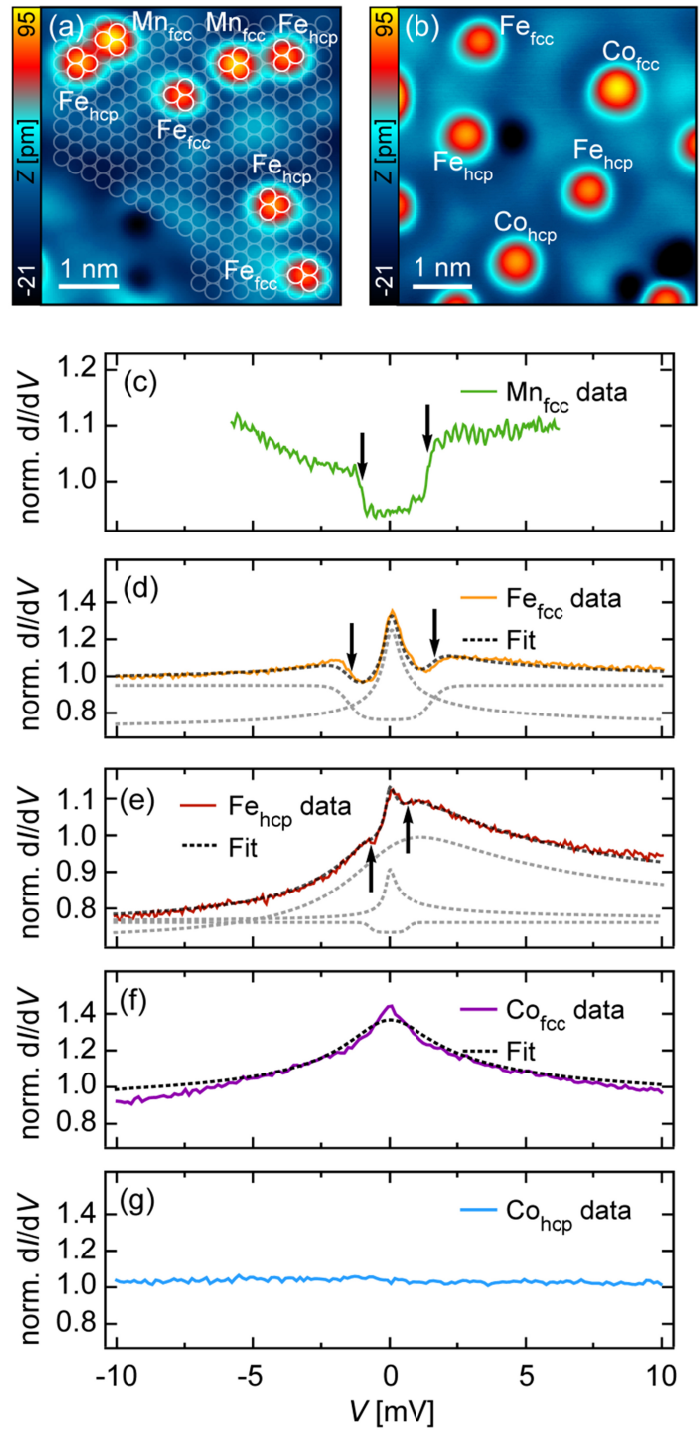

**Figure 1 | Topographic and spectroscopic signatures of the 3d transition metal atoms. a** Constant-current STM image of Mn and Fe atoms on different adsorption sites as indicated. The lattice of surface Re atoms determined from atom-manipulation images is overlayed (white circles). **b** Constant-current STM image of Co and Fe atoms on different adsorption sites as indicated ($V$ = 6 mV, $I$ = 200 pA). **c-g** Tunnel spectra taken in the normal metal state of the substrate ($B_z = 50$ mT). Normalized tunnel spectra taken on (**c**) Mn$_{fcc}$ ($V_{stab}$ = -6 mV, $I_{stab}$ = 3 nA, $V_{mod}$ = 100 µV), (**d**) Fe$_{fcc}$, (**e**) Fe$_{hcp}$, (**f**) Co$_{fcc}$, and (**g**) Co$_{hcp}$ ($V_{stab}$ = 20 mV, $I_{stab}$ = 3 nA, $V_{mod}$ = 40 µV). For Fe, the black dashed lines are fits to the sum of Frota functions, and broadened step functions at the positions indicated by the black arrows (see the different contributions indicated by gray dashed lines shifted vertically for clarity). For Co$_{fcc}$, the black dashed line is a fit to a Frota function.

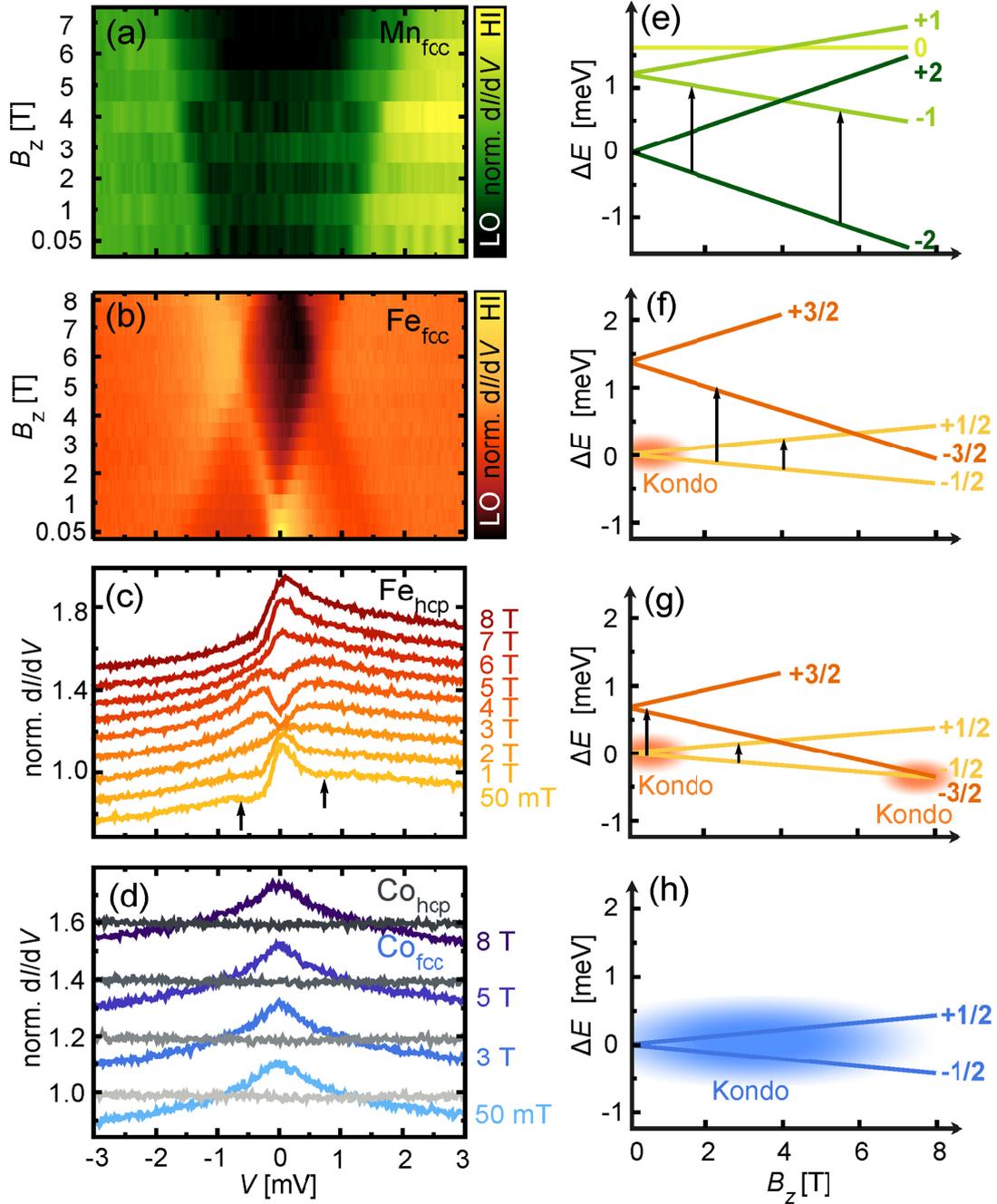

**Figure 2 | Magnetic-field-dependent tunnel spectra and energy level diagrams. a** Normalized tunnel spectra measured on $Mn_{fcc}$ ($V_{stab}$ = -6 mV, $I_{stab}$ = 3 nA, $V_{mod}$ = 100 μV, spectra from 0.05 to 3 T and 4 to 7 T were measured on two different atoms). **b** Normalized tunnel spectra measured on $Fe_{fcc}$ ($V_{stab}$ = 3 mV, $I_{stab}$ = 1 nA, $V_{mod}$ = 60 μV). **c** Normalized tunnel spectra measured on $Fe_{hcp}$ ($V_{stab}$ = 3 mV, $I_{stab}$ = 1 nA, $V_{mod}$ = 60 μV). **d** Normalized tunnel spectra measured on $Co_{hcp}$ (gray) and $Co_{fcc}$ (blue-purple, $V_{stab}$ = 3 mV, $I_{stab}$ = 1 nA, $V_{mod}$ = 60 μV). The spectra in (**c**) and (**d**) are vertically offset for clarity. The black arrows in (**c**) mark the position of the supposed inelastic excitation at $B_z$ = 50 mT. **e-h** Magnetic-field-dependent eigenenergies of the effective spin Hamiltonians which are consistent with the experimental data in the same row on the left, using the following parameters: (**e**) $Mn_{fcc}$, $S = 2$, $K = -1.56$ meV, $g = 2$; (**f**) $Fe_{fcc}$, $S = 3/2$, $K = +1.55$ meV, $g = 2$; (**g**) $Fe_{hcp}$, $S = 3/2$, $K = +0.79$ meV, $g = 1.5$; (**h**) $Co_{fcc}$, $S = 1/2$. The black arrows indicate the tunnel-electron-induced transitions from the ground state fulfilling the selection rules. The clouds illustrate the regions where many-body Kondo correlations can occur involving almost degenerate ground states separated by $\Delta S_z = +1$ or $-1$.

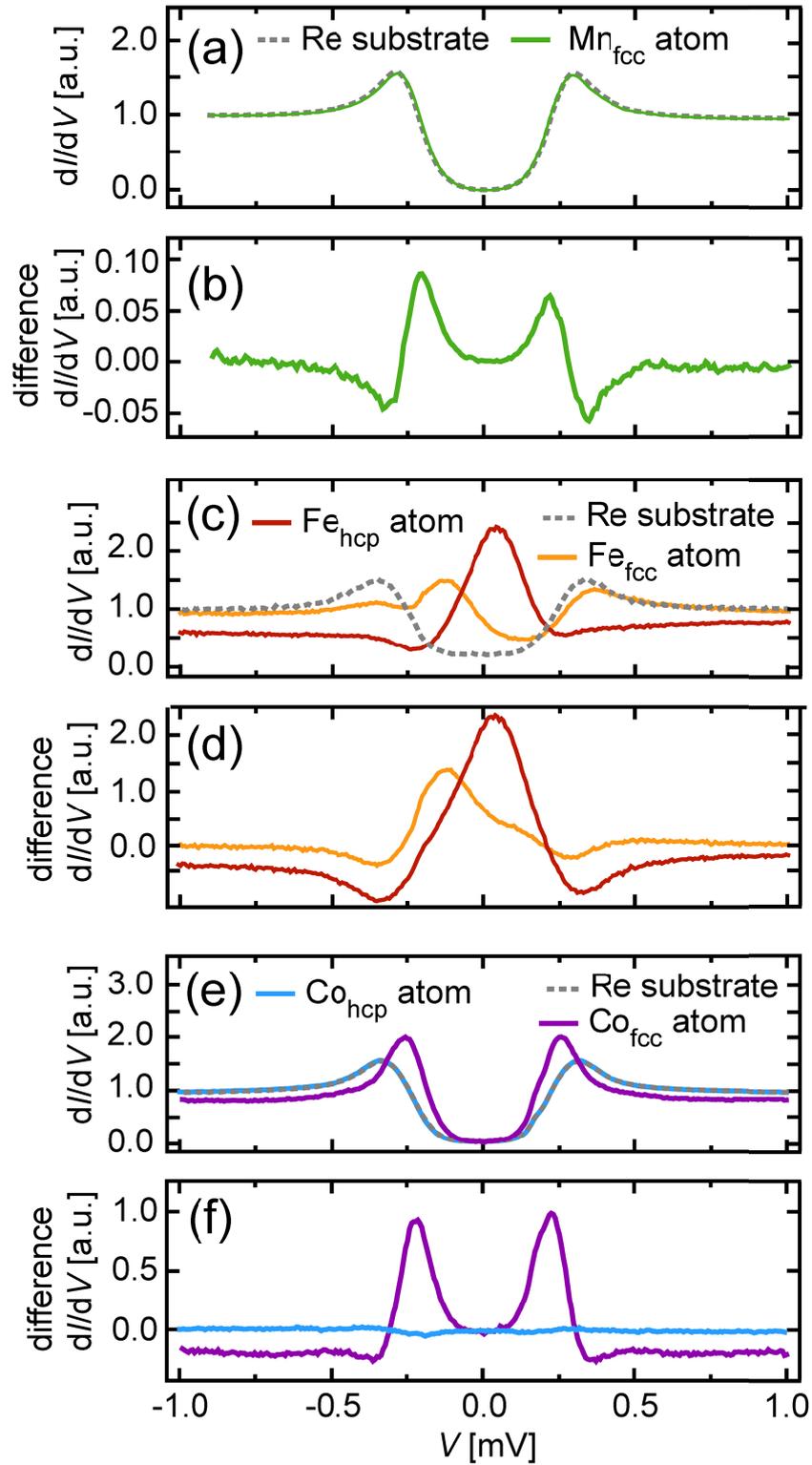

**Figure 3 | Yu-Shiba-Rusinov states. a, c, e** Raw tunnel spectra measured on (**a**) $Mn_{fcc}$ ($V_{stab}$ = -1 mV, $I_{stab}$ = 3 nA, $V_{mod}$ = 40 μV), (**c**) $Fe_{hcp}$ and $Fe_{fcc}$ ($V_{stab}$ = 1 mV, $I_{stab}$ = 0.2 nA, $V_{mod}$ = 40 μV), and (**e**) $Co_{hcp}$ and $Co_{fcc}$ atoms ($V_{stab}$ = 1 mV, $I_{stab}$ = 0.2 nA, $V_{mod}$ = 40 μV). The dashed lines are the substrate spectra taken for each species using the same microtip. All spectra have been measured at $B_z = 0$ T. **b, d, f** Normalized tunnel spectra from each of the raw spectra in the panel above, obtained by subtraction of the respective substrate spectrum.

# Supplementary Information
# Magnetism and in-gap states of 3d transition metal atoms on superconducting Re

Schneider *et al.*

## Supplementary Note 1 | Coexistence of YSR states with zero-bias resonances or spin excitations

For some of the investigated atom species, YSR states inside the energy gap of the substrate and one of the two other types of spectral features, i.e. zero-bias resonances or spin excitations, can be simultaneously observed in the same measurement, as shown in Supplementary Fig. 1. On the $Fe_{hcp}$ adatom, the YSR peak at $E \simeq E_F$ is observed simultaneously with the wider zero-bias resonance, where the latter is visible as a broad gentle maximum in the spectrum between $V = -3\,\mathrm{mV}$ and $V = +3\,\mathrm{mV}$. On the $Fe_{fcc}$ atom, the YSR states are observed simultaneously with the step-shaped spin excitations at $V = \pm 1.5\,\mathrm{mV}$. On the other hand, for both adsorption sites the narrow zero-bias features associated with the weak-coupling Kondo resonance shown in Fig. 1 of the main manuscript, as well as with the low-energy spin excitations of the $Fe_{hcp}$ adatom, are completely suppressed in the superconducting state of the Re substrate.

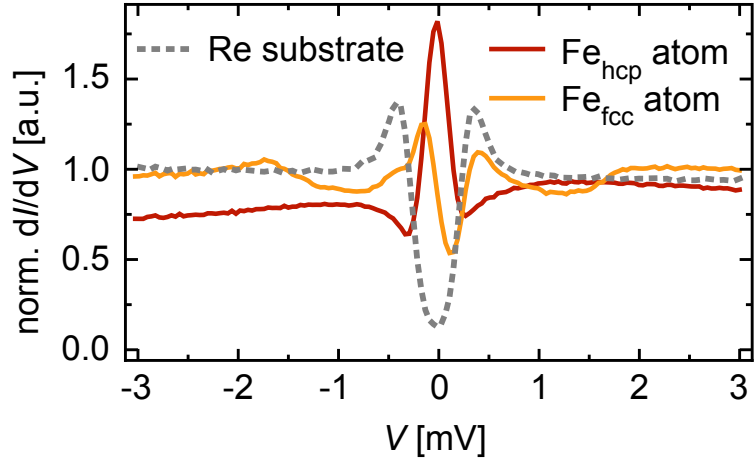

**Supplementary Figure 1 | Coexistence of spectral features.** Spectra on Fe atoms in the superconducting state of the Re substrate ($B_z = 0\,\mathrm{T}$), showing the coexistence of YSR states with wider zero-bias resonances or spin excitations ($V_{\mathrm{stab}} = 3\,\mathrm{mV}$, $I_{\mathrm{stab}} = 1\,\mathrm{nA}$, $V_{\mathrm{mod}} = 60\,\mu\mathrm{V}$).

## Supplementary Note 2 | Magnetic-field-induced reemergence of the Kondo resonance

In addition to the data of Fig. 2b in the main manuscript, we measured spectra on $Fe_{fcc}$ atoms in magnetic fields up to 11.5 T, where the crossing of the $S_z = -3/2$ and the $S_z = -1/2$ levels is expected from the effective spin model simulations (Fig. 2f and Supplementary Note 3). These spectra are shown in Supplementary Fig. 2. A broad resonance reemerges at this magnetic field, which can be interpreted as the Kondo resonance induced by the degeneracy of the two levels which fulfill the requirement of $\Delta S_z = \pm 1$ [1]. This observation further supports the level scheme of Fig. 2f.

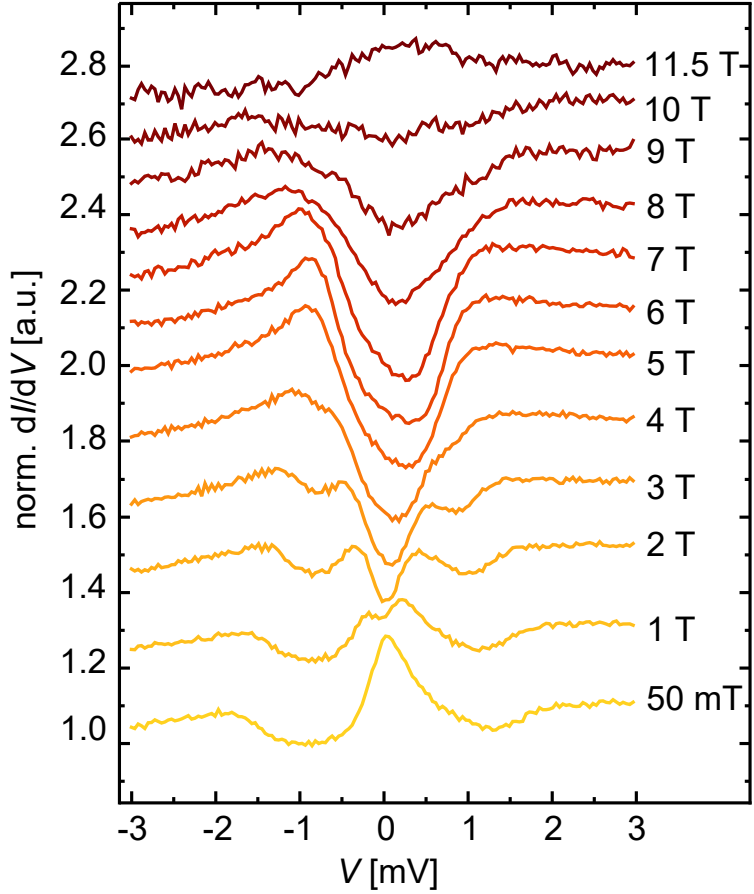

**Supplementary Figure 2 | Reemergence of a Kondo resonance on** $Fe_{fcc}$**.** STS spectra of an $Fe_{fcc}$ atom in external magnetic fields up to $11.5\,T$ ($V_{stab} = 3\,mV$, $I_{stab} = 1\,nA$, $V_{mod} = 60\,\mu V$).

**Supplementary Note 3 |**  **Fits and simulations of the experimental spectra**

In order to extract the relevant energy scales of the experimentally observed phenomena and compare to the *ab initio* calculations, the experimental spectra have been fitted using three different models which are each valid for different regimes of the external experimental and inherent parameters. The fitting and simulation procedures will be described in the following.

**Frota fit for** $B_z = 0$ **data of** $Fe_{fcc}$**,** $Fe_{hcp}$**, and** $Co_{fcc}$  In order to estimate the Kondo energy scale from the zero-bias features observed at zero magnetic field in the tunnel spectra of $Fe_{fcc}$, $Fe_{hcp}$, and $Co_{fcc}$ shown in Fig. 1 of the main manuscript, we used a single Frota function for $Fe_{fcc}$ and $Co_{fcc}$, a sum of two Frota functions for $Fe_{hcp}$, plus broadened step functions for $Fe_{fcc}$ and $Fe_{hcp}$ which capture the spin excitation:

$$dI/dV(V) = N_0 + \sum_{n=1}^{2} A_n Re \left[ e^{i\phi_n} \sqrt{\frac{i\Gamma_n}{i\Gamma_n + eV}} \right] \quad (1)$$
$$+ A_3 \left[ 1/(e^{\frac{eV+E_a}{\sigma}} + 1) - 1/(e^{\frac{eV-E_a}{\sigma}} + 1) \right].$$

Here, $N_0$ denotes the normal conductance, $A_n$ the relative intensities of the different features, $Re$ the real part, $\Gamma_n$ and $\phi_n$ the widths and parameters of the Frota functions, and $E_a$ and $\sigma$ the energy and width of the spin excitation. The fitted parameters are listed in Supplementary Table 1.

Using the half width at half maximum (HWHM) values of the Frota functions $\Gamma_{\mathrm{HWHM}} = 2.54\Gamma_{\mathrm{Frota}}$ one can estimate $k_B T_K = 0.687\Gamma_{\mathrm{Frota}}$, following Wilson's definition of the Kondo temperature [2,3]. The corresponding values for $k_B T_K$ are listed in Table 1 of the main manuscript. Note that for $Fe_{\mathrm{hcp}}$, the narrower of the two resonances ($\Gamma_2$) has been selected for this purpose. For $Mn_{\mathrm{fcc}}$ the spectra do not reveal any zero-bias resonance. Therefore, we assume that for this case, $k_B T_K$ is considerably smaller than the experimental thermal energy of $0.025\,\mathrm{meV}$.

|  | $Fe_{\mathrm{fcc}}$ | $Fe_{\mathrm{hcp}}$ | $Co_{\mathrm{fcc}}$ |
|---|---|---|---|
| $N_0$ | 0.892 | 0.748 | 0.846 |
| $A_1$ | 0.496 | 0.33 | 0.45 |
| $A_2$ | 0 | 0.146 | 0 |
| $A_3$ | 0.164 | 0.026 | 0 |
| $\Gamma_1$ [meV] | 0.264 | 2.29 | 1.50 |
| $\Gamma_2$ [meV] | - | 0.147 | - |
| $\phi_1$ | -0.192 | -0.7 | -0.10 |
| $\phi_2$ | - | -0.415 | - |
| $\sigma$ [meV] | 0.24 | 0.1 | - |
| $E_a$ [meV] | 1.52 | 0.7 | - |

**Supplementary Table 1 | Fit parameters for the Frota fits.** Fit parameters for the $Fe_{\mathrm{fcc}}$, $Fe_{\mathrm{hcp}}$ and $Co_{\mathrm{fcc}}$ spectra in Fig. 1 of the main manuscript using Eq. (1). Since the values for d$I$/d$V$ are in arb. units, all scaling parameters are chosen to be unitless.

**Perturbation-theory model for magnetic-field-dependent data of** $\mathrm{Mn_{fcc}}$ **and** $\mathrm{Fe_{fcc}}$  In order to extract the magnetic anisotropies from the magnetic-field-dependent spectra of $\mathrm{Mn_{fcc}}$ and $\mathrm{Fe_{fcc}}$, the spectra were simulated by using a third-order perturbative model [4]. The best fit to the experimental data is found for the parameters in Supplementary Table 2

Here, $S$, $g$ and $K$ are the parameters of the effective spin Hamiltonian, given in the main manuscript, $U$ is the potential scattering parameter, $J_\mathrm{K} \cdot \rho_\mathrm{F}$ is the product of the sample's electron density of states at the Fermi level and the Kondo coupling constant, and $T_\mathrm{eff}$ is the effective temperature. The results of the corresponding simulations are shown in Supplementary Figs. 3c and d, which are found to be in excellent agreement with the experimental data in Supplementary Figs. 3a and b.

|  | $S$ | $g$ | $K(= D \cdot S^2)$ [meV] | $U$ | $J_\mathrm{K} \cdot \rho_\mathrm{F}$ | $T_\mathrm{eff}$ [K] |
|---|---|---|---|---|---|---|
| $\mathrm{Mn_{fcc}}$ | 2 | 2 | -1.56 | 0 | -0.02 | 1 |
| $\mathrm{Fe_{fcc}}$ | 3/2 | 2 | +1.55 | 0.44 | -0.45 | 0.7 |

**Supplementary Table 2 | Fit parameters for the perturbation-theory simulation.** Fit parameters from the third-order perturbation-theory model used for the $\mathrm{Mn_{fcc}}$ and $\mathrm{Fe_{fcc}}$ spectra in Supplementary Figs. 3c and d, as well as for the energy level diagrams in Figs. 2e and f of the main manuscritpt.

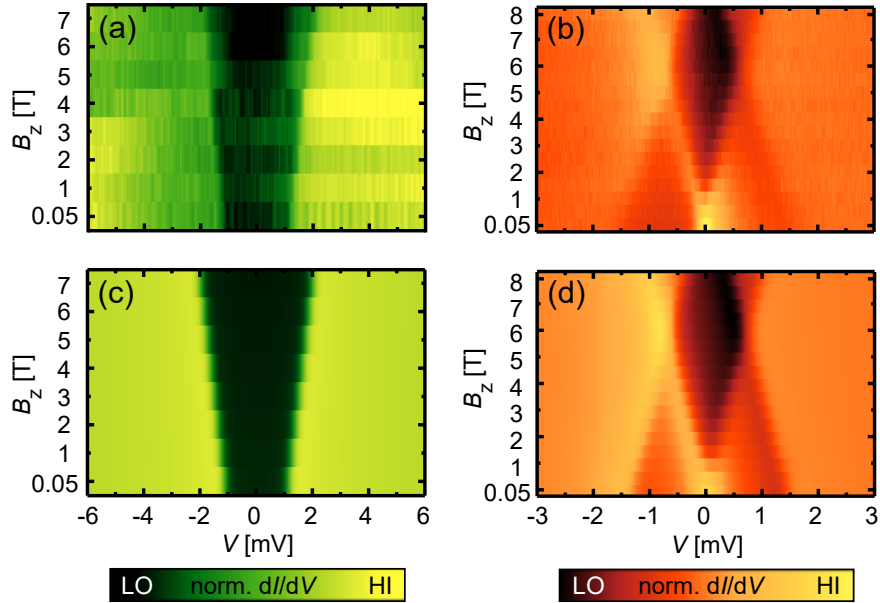

**Supplementary Figure 3 | Comparison of experimental data and perturbation-theory simulation.** (a) Experimental spectra measured on a $\mathrm{Mn_{fcc}}$ atom, taken from Fig. 2a of the main manuscript. (b) Experimental spectra measured on a $\mathrm{Fe_{fcc}}$ atom, taken from Fig. 2b of the main manuscript. (c) and (d) Perturbation-theory model simulations of the experimental spectra, using the parameters specified in the text.

**Multi-Frota model for magnetic-field-dependent data of** $\text{Fe}_{\text{hcp}}$ The magnetic-field-dependent spectra measured on $\text{Fe}_{\text{hcp}}$ have a relatively asymmetric Fano line shape which is most probably resulting from the quantum interference of the two tunneling channels, one into the hybridized $d$-states of the atom and one into the continuum of conduction-band states of the substrate [5]. Fano line shapes are not yet included in the perturbation-theory code [4]. However, we found that we can simulate the spectra reasonably well using a superposition of three Frota functions, and a broadened step function:

$$\mathrm{d}I/\mathrm{d}V(V) = N_0 + \sum_{n=1}^{3} A_n Re\left[e^{i\phi_n}\sqrt{\frac{i\Gamma_n}{i\Gamma_n + eV - E_n}}\right] \\ + A_4\left[1/(e^{\frac{eV+E_a}{\sigma}}+1) - 1/(e^{\frac{eV-E_a}{\sigma}}+1)\right]. \quad (2)$$

Here, $N_0$ denotes the normal conductance, $A_n$ the relative intensities of the different features, $Re$ the real part, $E_n$, $\Gamma_n$, and $\phi_n$ the centers, widths and parameters of the Frota functions, and $E_a$ and $\sigma$ the energy and width of the spin excitation. The fitted functions are plotted together with the experimental spectra in the left panel of Supplementary Fig. 4, while the different contributions are plotted in the right panel. The fitted parameters are collected in Supplementary Table 3. The first contribution, which is the wider resonance at the Fermi energy ($E_1 = 0\,\text{meV}$, gray dashed line in the right panel) is nearly magnetic-field independent and has a width of $\Gamma_1 = 2.29\,\text{meV}$ and $\phi_1 = -0.7$. In contrast, the narrower resonance splits up into two resonances ($E_2$ and $E_3$, black solid lines in the right panel) in the magnetic field. These resonance are at nearly symmetric energies on both sides of the Fermi level ($E_2 \approx -E_3$, $\Gamma_2 \approx \Gamma_3$, and $\phi_2 \approx \phi_3$), with an energy that linearly increases with $B_z$ up to about $4\,\text{T}$ and then decreases linearly with a Fermi level crossing at about $8\,\text{T}$ where the two resonances merge and split again. This energetic dependence of $E_2$ can be used to determine the parameters of the effective spin Hamiltonian, i.e. the $g$-factor and the magnetic anisotropy $K$. It can be rationalized assuming a $S = 3/2$ system with easy-plane anisotropy (see Fig. 2g of the main manuscript) by interpreting the narrow resonances as the splitting Kondo resonances from (i) the degenerate $S_z = \pm 1/2$ ground states for $B_z < 4\,\text{T}$, and (ii) from the crossing $S_z = -3/2$ and $S_z = -1/2$ levels for $B_z > 4\,\text{T}$. The additional broadened step functions at $E_a$ (orange solid lines) can be assigned to the spin excitation from the $S_z = -1/2$ to the $S_z = -3/2$ level, which coexists with the splitting Frota function for $B_z < 8\,\text{T}$, and then vanishes as the $S_z = -3/2$ level becomes the ground state. Indeed, the data can be fitted reasonably well, as shown in the left panel of Supplementary Fig. 4, by the following procedure. Initially, values for the parameters $K$ and $g$ were guessed. Then, using these values, the $B_z$ dependent excitation energies $E_2 = -E_3$ and $E_a$ were calculated from the effective spin Hamiltonian, and inserted into Eq. (2) to calculate the magnetic-field-dependent spectra. These were adjusted to the experimental spectra by variation of the remaining parameters of Eq. (2). Then, the procedure has been reiterated until we found the values $K = +0.79\,\text{meV}$ and $g = 1.5$ which best fit the experimental data and are used to plot the energy level diagram in Fig. 2g of the main manuscript.

| $B_z$ [T] | $N_0$ | $A_1$ | $A_2$ | $A_3$ | $E_2$ [meV] |
|---|---|---|---|---|---|
| 0.05 | 0.700 | 0.287 | 0.115 | 0.115 | 0.00 |
| 1 | 0.682 | 0.322 | 0.111 | 0.088 | 0.09 |
| 2 | 0.636 | 0.388 | 0.070 | 0.076 | 0.17 |
| 3 | 0.630 | 0.389 | 0.065 | 0.159 | 0.26 |
| 4 | 0.642 | 0.364 | 0.089 | 0.182 | 0.35 |
| 5 | 0.628 | 0.362 | 0.120 | 0.186 | 0.26 |
| 6 | 0.652 | 0.331 | 0.071 | 0.113 | 0.17 |
| 7 | 0.602 | 0.331 | 0.141 | 0.152 | 0.09 |
| 8 | 0.598 | 0.280 | 0.163 | 0.163 | 0.00 |
| 9 | 0.586 | 0.294 | 0.163 | 0.122 | 0.09 |
| 10 | 0.602 | 0.268 | 0.182 | 0.137 | 0.17 |
| 11.5 | 0.616 | 0.281 | 0.168 | 0.113 | 0.30 |

| $B_z$ [T] | $\phi_2$ | $\Gamma_2$ [meV] | $A_4$ | $E_a$ [meV] | $\sigma$ [meV] |
|---|---|---|---|---|---|
| 0.05 | -0.63 | 0.100 | 0.029 | 0.70 | 0.100 |
| 1 | -0.45 | 0.100 | 0.029 | 0.61 | 0.100 |
| 2 | -0.24 | 0.197 | 0.059 | 0.52 | 0.143 |
| 3 | -0.13 | 0.142 | 0.105 | 0.44 | 0.100 |
| 4 | -0.24 | 0.112 | 0.149 | 0.35 | 0.117 |
| 5 | -0.22 | 0.154 | 0.125 | 0.44 | 0.119 |
| 6 | -0.62 | 0.253 | 0.021 | 0.52 | 0.177 |
| 7 | -0.55 | 0.227 | 0.009 | 0.61 | 0.178 |
| 8 | -0.63 | 0.325 | 0.017 | 0.70 | 0.178 |
| 9 | -0.62 | 0.325 | - | - | - |
| 10 | -0.67 | 0.300 | - | - | - |
| 11.5 | -0.55 | 0.350 | - | - | - |

**Supplementary Table 3** | **Fit parameters for the** $Fe_{hcp}$ **spectra.** Fit parameters for the $Fe_{hcp}$ spectra in Supplementary Fig. 4 using Eq. (2). Since the values for d$I$/d$U$ are in arb. units, all scaling parameters are chosen to be unitless. The spin excitation intensity $A_4$ must vanish for $B_z > 8$ T, thus the parameters $\sigma$ and $E_a$ are undefined.

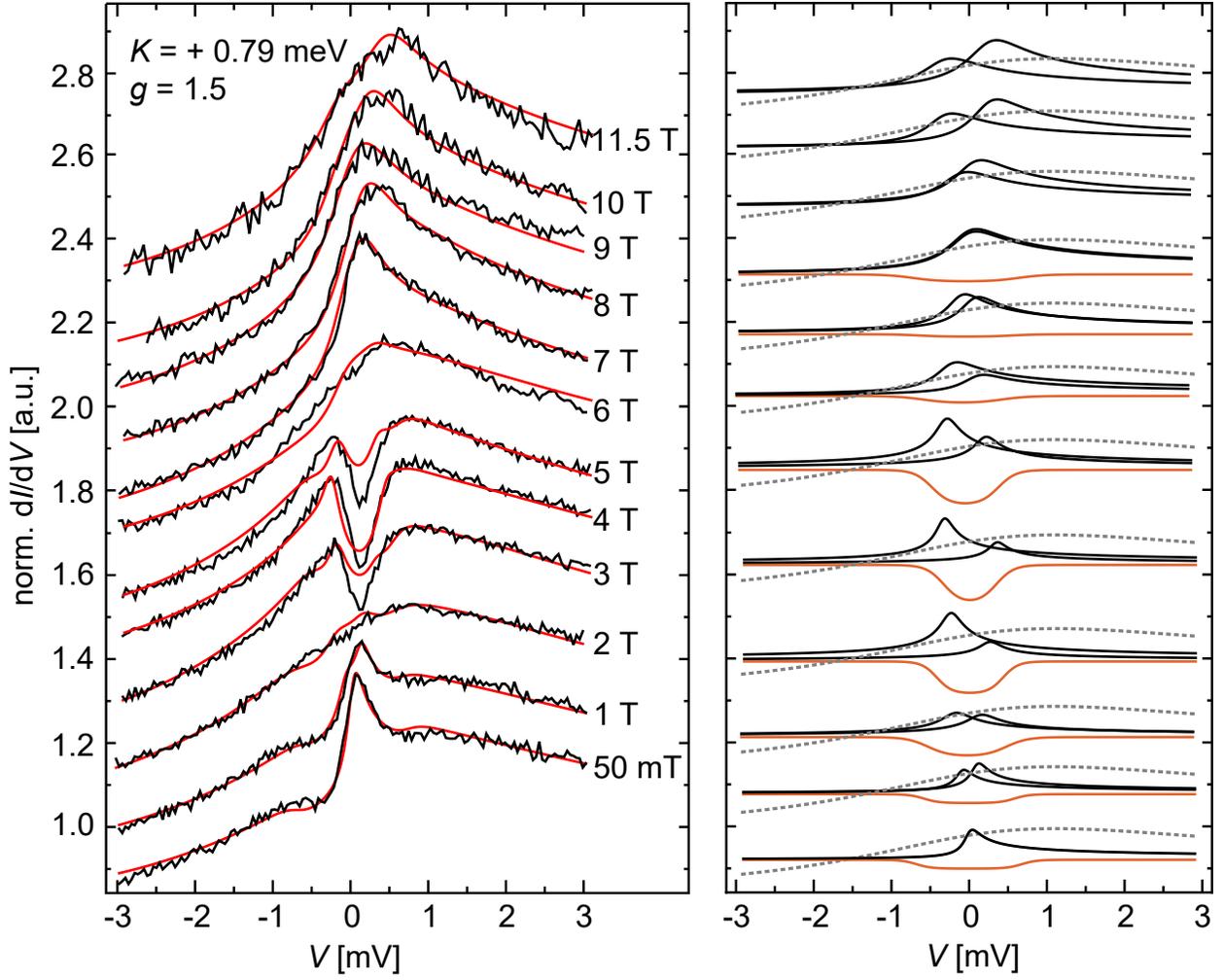

**Supplementary Figure 4 | Multi-Frota fit of the** $Fe_{hcp}$ **spectra.** Left: STS spectra of an $Fe_{hcp}$ atom in the external magnetic fields as indicated (black) and the corresponding fits (red). Right: The individual contributions to the fits given on the left side. We used a linear combination of a wider Frota function (gray dashed lines), two narrower Frota functions (black solid lines), and a broadened step function (orange solid lines).

**Supplementary Note 4 | Relaxed geometries obtained from the VASP method**

The interlayer distances in the KKRJ and KKRBp calculations were based on the geometries optimized by the VASP method, where in each considered case the $3d$ adatom pulled the neighboring three Re atoms slightly upwards compared to the other atoms in the top layer. Since during the KKR calculations each atom in the same layer has to occupy the same vertical position, the distance between the top Re layer and the one directly below it was determined from the average vertical position of all $49$ Re atoms in the top layer, while for the distance between the adatom and the surface only the three neighboring Re atoms were taken into account. The interlayer distances determined using this procedure are summarized in Supplementary Table 4.

|  | Re1-Re2 $d$ (Å) | adatom-Re1 $d$ (Å) |
|---|---:|---:|
| $Mn_{fcc}$ | 2.114 | 2.053 |
| $Mn_{hcp}$ | 2.114 | 2.000 |
| $Fe_{fcc}$ | 2.114 | 1.908 |
| $Fe_{hcp}$ | 2.114 | 1.849 |
| $Co_{fcc}$ | 2.114 | 1.866 |
| $Co_{hcp}$ | 2.115 | 1.746 |

**Supplementary Table 4 | Interlayer distances.** Interlayer distances for the considered adatoms in both possible adsorption sites, obtained from VASP calculations.

**Supplementary Note 5 | Calculated magnetic moments**

The magnitudes of the magnetic moments obtained from the three different theoretical methods, namely VASP, KKRJ, and KKRBp, are compared in Supplementary Table 5. Note that the KKRJ values have been obtained within the full-potential (FP) description, which allows an accurate treatment of the atomic cell geometry, instead of the atomic sphere approximation (ASA) method. The latter was used in the context of time-dependent density functional theory (TD-DFT) to obtain the Kondo coupling parameters listed in the main text. Spin $\mu_s$ and orbital $\mu_o$ contributions to the total magnetic moments $\mu$ are also listed in Supplementary Table 5. A negative sign of $\mu_o$ indicates that the orbital moment is pointing oppositely to the spin moment. The orbital moments are at least one order of magnitude smaller than the spin moments. In the case of Mn with a half-filled $3d$ band the almost complete disappearance of the orbital moment is expected based on Hund's rules, while for Fe and Co the small values of $\mu_o$ can be attributed to the crystal field quenching of the orbital moments. The decreasing trend of $\mu_s$ with increasing filling of the $3d$ band from Mn through Fe to Co may also be explained based on Hund's rules. The difference in $\mu$ between the adatom and that of the complete supercell in VASP or the cluster in KKR is caused by the induced moments in the Re substrate; however, these induced moments are significantly smaller than in more easily polarizable substrates such as Pt or Pd [6]. For the same species, the magnetic moments in the hcp adsorption sites are consistently smaller than in the fcc sites. As shown in Supplementary Table 4, it was found within the VASP calculations that the adatoms in the hcp sites always relax closer to the substrate than in the fcc site, which results in an enhanced hybridization of the adatom's electronic states with the wide 5$d$ band of Re and a decreased magnetic moment.

| adatom | VASP | | | KKRJ FP | | | KKRBp ASA | | |
|---|---|---|---|---|---|---|---|---|---|
| | $\mu_s$ | $\mu_o$ | $\mu$ | $\mu_s$ | $\mu_o$ | $\mu$ | $\mu_s$ | $\mu_o$ | $\mu$ |
| $Mn_{fcc}$ | 3.91 | 0.02 | 3.94 | 3.99 | 0.03 | 4.02 | 3.74 | 0.05 | 3.79 |
| $Mn_{hcp}$ | 3.74 | 0.02 | 3.76 | 3.70 | 0.05 | 3.75 | 3.42 | 0.02 | 3.44 |
| $Fe_{fcc}$ | 2.89 | 0.19 | 3.08 | 2.85 | 0.14 | 2.99 | 2.60 | 0.21 | 2.81 |
| $Fe_{hcp}$ | 2.66 | 0.19 | 2.84 | 2.54 | 0.18 | 2.72 | 2.07 | 0.14 | 2.21 |
| $Co_{fcc}$ | 1.79 | 0.23 | 2.02 | 1.67 | 0.11 | 1.78 | 1.42 | 0.11 | 1.53 |
| $Co_{hcp}$ | 1.37 | 0.17 | 1.54 | 1.02 | 0.11 | 1.12 | 0.01 | 0.00 | 0.01 |

| cluster/ supercell | VASP | | | KKRJ FP | | | KKRBp ASA | | |
|---|---|---|---|---|---|---|---|---|---|
| | $\mu_s$ | $\mu_o$ | $\mu$ | $\mu_s$ | $\mu_o$ | $\mu$ | $\mu_s$ | $\mu_o$ | $\mu$ |
| $Mn_{fcc}$ | 4.22 | 0.04 | 4.26 | 4.17 | 0.02 | 4.19 | 4.36 | 0.26 | 4.62 |
| $Mn_{hcp}$ | 3.98 | 0.04 | 4.01 | 3.61 | -0.02 | 3.60 | 4.22 | 0.24 | 4.47 |
| $Fe_{fcc}$ | 3.32 | 0.22 | 3.54 | 3.03 | 0.10 | 3.13 | 3.17 | 0.48 | 3.64 |
| $Fe_{hcp}$ | 2.89 | 0.20 | 3.08 | 2.53 | 0.12 | 2.65 | 2.70 | 0.37 | 3.08 |
| $Co_{fcc}$ | 2.39 | 0.26 | 2.65 | 1.94 | 0.09 | 2.03 | 1.87 | 0.30 | 2.18 |
| $Co_{hcp}$ | 1.78 | 0.18 | 1.96 | 1.11 | 0.09 | 1.20 | 0.01 | 0.00 | 0.01 |

**Supplementary Table 5 | Calculated magnetic moments of the 3$d$ adatoms on the Re(0001) surface.** Spin $\mu_s$, orbital $\mu_o$ and total $\mu$ magnetic moments are listed in $\mu_B$ units, obtained from different methods: VASP with supercell calculations; KKRJ with the embedded cluster method within the FP treatment; and KKRBp with the embedded cluster method within the ASA. The values for just the 3$d$ adatom and for the complete system including induced moments are listed separately.

**Stoner model** The scanning tunneling spectroscopy data indicate no sign of magnetism for the $Co_{hcp}$ adatom, which is reproduced by the almost vanishing magnetic moment of the KKRBp method, while the other two methods yield a small but finite magnetic moment, as shown in Supplementary Table 5. The $Co_{fcc}$ adatom, in contrast to the hcp case, carries a rather large magnetic moment. This difference between the two adsorption sites of Co can be understood within the Stoner model. Magnetism is favored if the Stoner criterion is satisfied, i.e. $I\rho(E_F) > 1$, where $I$ is the Stoner parameter and $\rho(E_F)$ is the density of states (DOS) at the Fermi energy in the nonmagnetic state. As displayed in Supplementary Fig. 5, the DOS at $E_F$ as obtained with the KKRBp method for $Co_{hcp}$ is about half of the value characterizing $Co_{fcc}$. It can be seen that while for the fcc adsorption site a peak is located at the Fermi energy, this is suppressed for the hcp site.

If one uses the Stoner parameter obtained by Janak [7] for bulk Co, $I = 0.49\,\text{eV}$, one obtains $I\rho(E_\text{F}) = 0.95$ for the hcp site and $I\rho(E_\text{F}) = 1.64$ for the fcc site. In other words, in contrast to $\text{Co}_\text{fcc}$, $\text{Co}_\text{hcp}$ does not satisfy the Stoner criterion and misses it by a small amount. This indicates that a somewhat larger $\rho(E_\text{F})$ could provide an explanation for the small magnetic moments obtained with the other two methods, VASP and KKRJ. We note that simulations with the KKRJ code considering a $\text{Co}_\text{hcp}$ atom with a slight additional relaxation towards the surface leads to the disappearance of its magnetic moment, similarly to the result of the KKRBp method presented here. As also shown in Supplementary Fig. 5, about $60\%$ of the difference in the DOS at the Fermi level between the two adsorption sites can be attributed to the $d_{xz}$ and $d_{yz}$ orbitals, which display a peak at the Fermi level for $\text{Co}_\text{fcc}$, but no peak for $\text{Co}_\text{hcp}$. This indicates that primarily these two orbitals are responsible for the different magnetic properties, which is reasonable given that these two orbitals overlap most strongly between the adatoms and the Re atoms in the layer below.

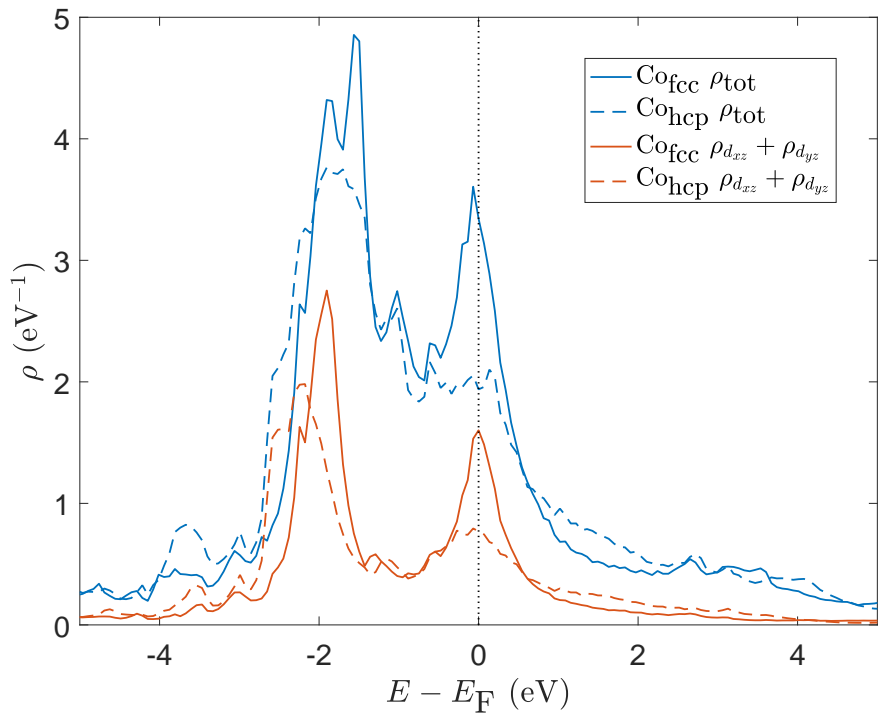

**Supplementary Figure 5 | Calculated DOS for the two adsorption sites of the Co adatom.** Nonmagnetic DOS for the two adsorption sites of Co obtained within the KKRBp method.

**Supplementary Note 6 | Calculation of the Kondo coupling**

To evaluate the effective magnetic exchange coupling of the spin of the adatom to the substrate electrons from first principles, the so-called s-d coupling or Kondo coupling constant, $J_K$, we first refer to the work of Schrieffer and Wolf [8].

Considering the Anderson impurity model [9] and assuming weak hybridization strength, $V$, between the electronic states of a single impurity and those of the bath delocalized electrons, $J_K$ simplifies to $2|V|^2 U/E_d^\uparrow E_d^\downarrow$, where $U$ corresponds to the Coulomb repulsion between opposite-spin $d$ electrons located at $E_d^\uparrow$ and $E_d^\downarrow$ energies. To obtain their results, Schrieffer and Wolf assumed a constant DOS for the bath electrons.

The spin-excitation spectra are heavily affected by the hybridization strength, $V$, which leads to their damping and determines the so-called effective Gilbert damping $\alpha$ [10]. Utilizing TD-DFT and linear response theory within the ASA [11,12], one can approximately relate the damping to the DOS at the Fermi energy of the excited impurity as follows [13]:

$$\alpha \simeq \frac{\gamma \mu_s U^2 \pi}{8} \rho_\uparrow(E_F) \rho_\downarrow(E_F), \quad (3)$$

where $\mu$ is the spin magnetic moment, $\rho_\uparrow$ and $\rho_\downarrow$ are the spin-resolved DOSs at the Fermi energy of the magnetic atom, and $\gamma$ is the gyromagnetic ratio, which differs from 2 due to the hybridization with the host atoms. $U$ is the exchange and correlation kernel of the impurity, leading to the exchange splitting of the impurity state [11].

Equation (3) is obtained after mapping the transversal magnetic response function obtained with TD-DFT to the one obtained from the classical Landau-Lifshitz-Gilbert (LLG) equation [10,14],

$$\frac{d\boldsymbol{\mu}_s}{dt} = -\gamma (\boldsymbol{\mu}_s \times \boldsymbol{B}) + \alpha \frac{\boldsymbol{\mu}_s}{\mu_s} \times \frac{d\boldsymbol{\mu}_s}{dt}. \quad (4)$$

After expressing the DOS given in Eq. (3) in terms of the hybridization function and making similar assumptions to those of Schrieffer and Wolf [8], one finds

$$\alpha = \frac{\gamma \mu_s \pi}{32} (J_K \cdot \rho_F)^2 \quad, \quad (5)$$

where $\rho_F$ is the DOS at the Fermi energy of the surface Re atoms in the absence of the impurity.

In Supplementary Table 6 the values of $\gamma$, $\alpha$, $\rho_F$, $|J_K|$, and $|J_K \cdot \rho_F|$ are listed, obtained for the different considered adatoms and stackings, respectively. The values of $|J_K \cdot \rho_F|$ are also given in the table of the main manuscript for comparison with the experimentally determined trend of the Kondo temperatures.

|  | $\mu_s$ | $\gamma$ | $\alpha$ | $\rho_F$ | $|J_K|$ | $|J_K \cdot \rho_F|$ |
|  | ($\mu_B$) | ($\mu_B/\hbar$) |  | (eV$^{-1}$) | (eV) |  |
|---|---|---|---|---|---|---|
| Mn$_{fcc}$ | 4.050 | 1.668 | 0.067 | 0.795 | 0.399 | 0.317 |
| Mn$_{hcp}$ | 3.882 | 1.640 | 0.111 | 0.782 | 0.538 | 0.421 |
| Fe$_{fcc}$ | 2.856 | 1.783 | 0.334 | 0.791 | 1.033 | 0.817 |
| Fe$_{hcp}$ | 2.596 | 1.938 | 0.408 | 0.748 | 1.215 | 0.909 |
| Co$_{fcc}$ | 1.439 | 2.746 | 0.893 | 0.814 | 1.864 | 1.517 |
| Co$_{hcp}$ | 0.863 | 4.815 | 1.576 | 0.797 | 2.466 | 1.965 |

**Supplementary Table 6 | Parameters used for the calculation of the Kondo coupling.** $\mu_s$ is the spin magnetic moment, $\gamma$ is the effective gyromagnetic ratio, $\alpha$ is the Gilbert damping, $\rho_F$ is the density of states at the Fermi level of the surface Re atoms in the absence of the impurity, and $|J_K|$ is the absolute value of the Kondo coupling constant. Note that the spin magnetic moments calculated within the ASA are slightly different from the ones obtained within the FP method listed in Supplementary Table 5.

## Supplementary References